# Optimising PICCANTE – an Open Source Particle-in-Cell Code for Advanced Simulations on Tier-0 Systems

A. Sgattoni[a], L. Fedeli[a,b], S. Sinigardi[a,c], A. Marocchino[d], A. Macchi[a,b*]
V. Weinberg[e†], A. Karmakar[e]

[a]*CNR/INO Via Moruzzi 1, 56124 Pisa, Italy*
[b]*University of Pisa, Largo Pontecorvo 3, 56127 Pisa Italy*
[c]*University of Bologna, Dipartimento di Fisica e Astronomia, Via Irnerio 46, 40126, Bologna, Italy*
[d]*"La Sapienza" - Università di Roma, Italy*
[e]*Leibniz Supercomputing Centre of the Bavarian Academy of Sciences and Humanities, 85748 Garching b. München, Germany*

**Abstract**

We present a detailed strong and weak scaling analysis of PICCANTE, an open source, massively parallel, fully-relativistic Particle-In-Cell (PIC) code. PIC codes are widely used in plasma physics and astrophysics to study the cases where kinetic effects are relevant. PICCANTE is primarily developed to study laser-plasma interaction. Within a PRACE Preparatory Access Project, various revisions of different routines of the code have been analysed on the HPC systems JUQUEEN at Jülich Supercomputing Centre (JSC), Germany, and FERMI at CINECA, Italy, to improve scalability and I/O performance of the application. The diagnostic tool Scalasca is used to identify suboptimal routines. Different output strategies are discussed. The detailed strong and weak scaling behaviour of the improved code are presented in comparison with the original version of the code.

1. **Introduction**

This whitepaper describes a detailed scaling analysis and various efforts to optimise the Particle-In-Cell (PIC) code PICCANTE for massively parallel simulations of plasma physics on Tier-0 systems. PICCANTE [1] has been recently developed as an Open Source project and an alpha version has been released. The code aims to allow a strong flexibility, versatility and readability in order to be suitable for a broad number of developers as well as users. While the final goal of the PICCANTE project is to provide a general tool for kinetic simulations of plasma physics, in the near term the code will be mainly used to simulate the interaction of plasmas with superintense laser pulses in various regimes and for different applications. In particular, we aim to make the code suitable as a tool for experimental physics, finding the best compromise between complexity and flexibility, and between physical realism and model simplicity. Applications of current interest include laser-plasma acceleration [2,3], high field plasmonics [4], laser-driven shock waves [5] and pair plasmas of astrophysical interest [6]. In particular, high field plasmonics simulations require the ability to simulate complex geometries for both target and irradiation, including micro- and nano-structuring of the target. For shock wave simulations it is essential to follow the laser interaction and the dynamics of the plasma over large spatial and temporal scales, which require very large computing resources and efficient algorithms.

The computations have been performed on the IBM BlueGene/Q systems JUQUEEN at Jülich Supercomputing Centre (JSC), Germany, and FERMI at CINECA, Italy. An IBM PowerPC A2 node of both JUQUEEN and FERMI has 16 cores available, clocked at 1.6 GHz. The peak performance is 204.8 GFlops/node. More information about the two systems can be found in a Best Practice Guide published by PRACE [7].

---

* Principal investigator, E-mail address: andrea.macchi@ino.it
† Coordinating PRACE expert, E-mail address: weinberg@lrz.de

## 2. Description of the Code and Typical Results

A PIC code provides an approximate Lagrangian solution of the Maxwell-Vlasov system of equations. The Vlasov equation for the phase space distribution function $f_a = f_a(r, p, t)$ of a species $a$ of charged particles with mass $m_a$ and charge $q_a$ in the presence of electromagnetic (EM) fields and without collisions is

$$\left(\partial_t + v \cdot \nabla_r + q_a(E + v \times B) \cdot \nabla_p\right) f_a(r, p, t) = 0,$$

where $v = p/(p^2 + m_a^2)^{1/2}$ (suitably normalised, dimensionless quantities are used everywhere). In the PIC approach, the following discrete approximation is introduced:

$$f_a(r, p, t) = A \sum_n g\left(r - r_n(t)\right) \delta(p - p_n(t)).$$

The time-dependent functions $r_n = r_n(t)$ and $p_n = p_n(t)$ are the trajectories of (macro-) particles sampling the phase space, which satisfy the equations of motion

$$dp_n/dt = q_a\left(E(r_n, t) + v_n \times B(r_n, t)\right), \qquad dr_n/dt = v_n.$$

The function $g(r)$ describes the spatial "shape" of the particle; it is compact and positive definite. The self-consistent coupling to EM fields occurs via the calculation of the current density,

$$J(r, t) = \sum_{a,n} q_a v_n g(r - r_n(t)),$$

and the fields are advanced in time using Maxwell's "curl" equations $\partial_t E = \nabla \times B - J$ and $\partial_t B = -\nabla \times E$. Both current density and EM fields are defined on a spatial grid. In order to calculate the force on a particle, a spatial average of the EM fields weighted with $g(r)$ is performed.

Currently, PICCANTE uses standard PIC algorithms. A leap-frog scheme and the "Boris pusher" algorithm (both with second order accuracy in time) are used to advance the particle trajectories [8] while a FDTD scheme on the Yee lattice [9] is used to advance the EM fields (details on enforcing the continuity equation, boundary conditions, et cetera are not given here for brevity). In the implementation for HPC systems MPI is used for the parallelisation of the C++ code. The simulation box, EM fields and particles, is divided into spatial sub-domains each assigned to a computing node. Data exchange between nodes includes both EM fields and particles propagating across different domains. The algorithm stability condition, the standard stability condition for the EM field solver (it ensures that no numerical divergence arises) ensures that $\Delta x_i \geqslant c\Delta t$, where $\Delta t$ is the time step, $c$ is the speed of light and $\Delta x_i$ is the cell size. Hence, the particles cannot cross more than a cell edge per time step and all data communication occurs only between nearest neighbours because both particles and fields do not propagate farther than one cell within one time step. In three-dimensional simulations of laser-plasma interactions recently published by our group [3], up to some tens of billions of grid points and particles have been used, with total memory requirements exceeding 1 Terabyte. Simulation output includes complete maps of the EM fields and phase space distribution of the particles. Such requirements necessitate the usage of Tier-0 systems and impose an efficient code parallelisation.

Fig. 1 presents some typical results of 3D simulations using PICCANTE. The left figure displays a result from a recent study of laser-driven ion acceleration considering innovative targets, e.g. foam attached thin foils. The figure shows the electron density of a "near-critical" density plasma having an electron density near the transparency threshold. Shown is a 3D view of the electron density in the z > 0 region corresponding to half the simulation box. The laser is injected at 30° from the bottom left side. The laser pulse drills a low-density channel. Electrons are swept and accelerated by the laser pulse forming high density bunches in the plasma (red regions) and being ejected from the top side (green arch-like structures). This particular configuration is designed to increase the conversion efficiency of laser-energy into electron kinetic energy and to enhance proton acceleration driven by space-charge electric field [10].

Another research topic for which PICCANTE is providing an essential insight is the interaction of a laser pulse with grating targets (e.g. solid foils with a period engraving on the irradiated surface). A typical result of this research is presented in Fig. 1 (b). The figure shows the electromagnetic field amplitude (red-blue tones) and the electron density of the dense target (grating in green). The laser pulse is incident from the top right side and undergoes diffraction by the grating. A sizeable fraction of the electromagnetic energy is absorbed as a surface wave travelling along the target (bottom left side). This activity aims at studying plasmonic effects in a nonlinear regime of high laser intensity so that the electrons dynamics is relativistic. This unexplored regime of high field plasmonics may open new ways to develop laser-driven sources [4] and to manipulate high power laser pulses on a sub-wavelength scale.

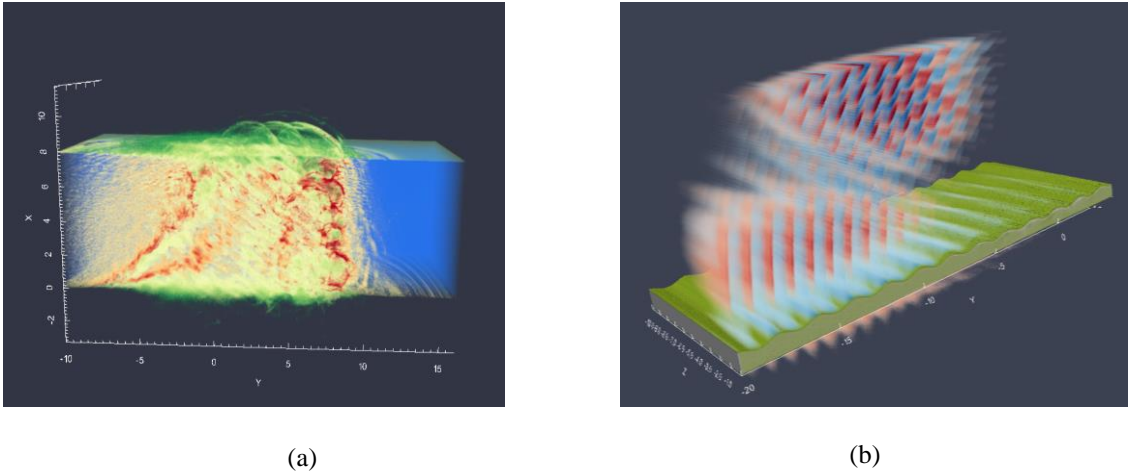

(a)                  (b)

Figure 1 – Typical results of two 3D simulations: Fig. (a) shows the electron density of a "near-critical" density plasma having an electron density near the transparency threshold. Fig. (b) presents the interaction of an ultraintense laser pulse with a grating target.

## 3. Code Optimisations

### 3.1 Improved Output Strategies

The output routines were extensively rewritten in order to allow greater flexibility and to reduce the use of disk space. The "output manager" now allows to produce reduced-dimension outputs (e.g. a plane in a 3D simulation or a 1D subset in a multidimensional run) and subdomain outputs (e.g. a small box in a large 3D simulation), as well as to control the output in time by choosing a suitable temporal window and frequency.

The old output strategy made use of *MPI_File_write*. For every output request (e.g. EM fields, particles' phase space, etc.), each MPI task had to write "simultaneously" in the same shared file. This method proved to be efficient for small numbers of MPI tasks (less than 1024), but became a critical bottleneck for larger number of MPI tasks on a BlueGene/Q machine, since concurrent writes are very inefficient to achieve a significant I/O bandwidth. Moreover, on the BlueGene/Q architecture, for every 128 compute nodes there is one I/O node, which has to manage all the multiple *MPI_File_write* requests coming from up to $128 \times 16 = 2048$ MPI tasks or even more if simultaneous multithreading is used.

Several parallel output strategies were developed and tested within this project. First, we switched to *MPI_File_write_all,* which improved the write times significantly, but still maintained a very bad scaling over 2048 MPI tasks. We also grouped the I/O within the MPI tasks so that each group had one task responsible for writing all the data of the group in the same big file. This improved the time, but not the scaling significantly. The bad scaling is due to the fact that the number of accesses to the same file is increasing proportionally to the total number of MPI tasks. Optimal scaling was achieved writing one single file per MPI task (task local I/O). However, it was not the fastest algorithm for small simulations (less than 2048 MPI tasks) and, more importantly, the extremely large amount of output files without proper directory hierarchy is a rather inefficient use of a parallel filesystem like GPFS. This strategy also made the post processing very hard and is absolutely unfeasible in this context.

Considering these aspects, eventually the following solution was adopted. Let $N$ be the total number of MPI tasks. We define groups of tasks of size $G$ and within each group the tasks send their data buffer to a single dedicated task, called the master task. Then, $M = N / G$ master tasks are writing data to files using the standard MPI I/O routine *MPI_File_write*. The number of output files is set to $F < M$, with $M / F$ master tasks writing data on each output file. This strategy is illustrated in the schematic of Fig. 2 (a), where we also show the old inefficient approach of all tasks writing to the same output file. For large simulations (e.g. $N > 2048$) the values we tested were in the 32-128 range for $G$ and $N / 1024$ or $N / 2048$ for $F$. This solution exploits the limited number of I/O nodes on the BlueGene/Q architecture and combines the advantages of very good scaling (as the number of accesses to the same file is kept constant for increasing total number of MPI tasks) and a reasonable number of output files. The new I/O strategy led to a major speedup of I/O, with a reduction of output time up to a factor 40 for particles' data and 600 for fields' data (see Fig. 2 (b) and Table 1), as well as a substantial improvement of the scaling. The reason for such large differences in the improvement of the particles' vs. fields' data depends on the average size of the files. The grid based data (e.g. densities and fields) for each MPI task have fixed size (the domain does not change during the simulation) and are based on a limited number of grid points. The particles' data, on the other hand, represent the phase space coordinates of the particles, whose number (per MPI task) may change in time and is about 50 times larger than for the fields in one of the test cases considered. The old output strategy required each MPI task to write data in the same file. Output data sizes per MPI task for the old method varied from 176 KB and 48 MB (for 512 MPI tasks) to 5.5 KB and 1.5 MB (for 16384 tasks, scaling extremely poorly) for the fields' and particles' data, respectively. Using the new output strategy, a writing performance of max. 5 GB/s was achieved when writing large files (16 files of 12 GB for each particle's output) on 16384 cores of JUQUEEN. Since on this particular machine each I/O node has a maximum bandwidth of 2 GB/s and is reserved for 2048 cores, the expected bandwidth using 16384 cores is 2 GB/s $\times$ 8 = 16 GB/s. PICCANTE is achieving approximately 31% of the maximum bandwidth and 50% of the average bandwidth.

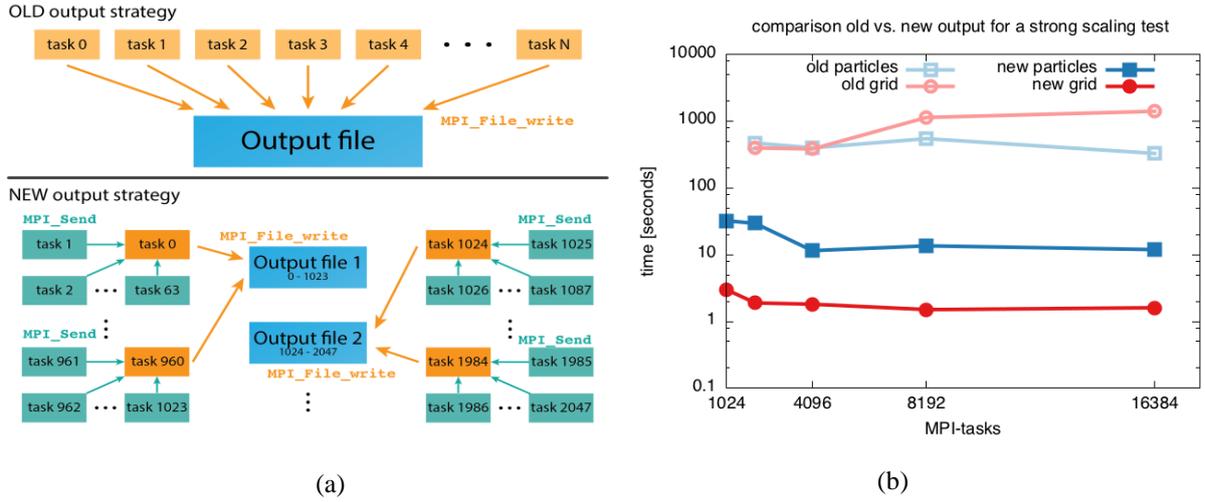

(a)  (b)

Figure 2 – (a) Schematic of the old and new output strategies in PICCANTE. In this example, $G = 64$ and $F = N/128$. (b) Comparison of the output times between the old (empty symbols in light shades of red and blue) and the new strategies (filled symbols in dark shades of red and blue) versus N. The time is given in seconds and corresponds to the aggregated time given by "Scalasca" divided by the number of MPI tasks N. The time spent to write the particles position is represented in red, while the total time spent for writing the grid based outputs (EM fields and densities) is represented in blue.

| Strong Scaling | Old Particles | New Particles | Old Grid | New Grid |
|---|---|---|---|---|
| 1024 |  | 32.1 |  | 2.98 |
| 2048 | 467.8 | 29.4 | 396.5 | 1.89 |
| 4096 | 397.9 | 11.5 | 382.3 | 1.81 |
| 8192 | 545.7 | 13.5 | 1132.8 | 1.50 |
| 16384 | 327.8 | 11.8 | 1403.8 | 1.60 |

(a)

| Weak Scaling | New Particles | New Grid |
|---|---|---|
| 512 | 16.4 | 1.72 |
| 1024 | 16.6 | 1.86 |
| 2048 | 29.4 | 1.89 |
| 4096 | 27.6 | 2.10 |
| 8192 | 46.5 | 2.73 |
| 16384 | 47.3 | 2.87 |

(b)

Table 1 – (a) Time (in seconds) required for writing the particles coordinates and the grid based data (EM fields and densities) in a strong scaling test comparing the new and old strategies. The same data is used for Fig. 2 (b). Table (b) reports the output times (in seconds) for a complete (from 512 to 16384 MPI tasks) weak scaling test which was feasible only using the new routines. The run for 2048 MPI tasks is the same for strong and weak scaling.

*3.2 Input-File Support*

PICCANTE is designed to exploit an object oriented structure: the user can "build" a simulation instantiating the various "elements": `particle`, `grid`, `em-field` etc. and then calling directly all the public functions needed to run the simulation itself. PICCANTE needed the user to set all the parameters of a simulation in a suitable "main.cpp" file, which was then compiled for each of the different runs. In the newest version, new functions have been introduced in order to initialise a simulation by reading a user-friendly JSON input file. A new main file was designed to allow typical simulation setup without any code editing. Support for JSON parsing is provided using the library "jsoncpp", which is licensed as "Public Domain" (or with MIT license in some jurisdictions) [12].

*3.3 Memory Allocation Strategies*

Particles' coordinates are stored in large arrays. Seven double precision variables per particle are stored: *(x, y, z, px, py, pz, w)* where *w* is the so called "weight" of a particle (use of variable weights allows for a more accurate resolution of low-density regions). We tested two main allocation strategies: *[x1, y1, ... , w1, x2, y2, ... , w2, x3 ...]* and *[x1, x2 , ... , xN, y1, y2 ,..., yN , z1 ...]*. On an Intel Linux cluster one of the strategies proved to be slightly better, but the difference was minimal and possibly problem-dependent. Since no relevant speedup was achieved on BlueGene/Q, no further efforts were spent on this topic.

4. **Scaling of the Code**

Strong and weak scaling measurements using a typical simulation case have been performed on both JUQUEEN and FERMI. The MPI implementation on the Blue Gene/Q system is derived from the MPICH2 implementation of the Mathematics and Computer Science Division (MCS) at Argonne National Laboratory. To test the code scalability we have chosen a standard test problem such as the two-stream instability: two counter-streaming beams of particles permeating the entire simulation box with periodic boundary conditions. The setup for the strong scaling tests consists of a simulation box of $512 \times 512 \times 128$ cells with 54 particles per cell. For the weak scaling analysis the same setup is used for 2048 cores. For larger core-counts the size of the box is changed accordingly to keep the problem size per processor constant.

The Scalasca utility [11] has been used to analyse the runtime behaviour of the code and break the total measured wall-clock time ("ALL") down into the time spent in MPI ("MPI"), purely local computations ("USR") and functions calling other functions ("COM"). The Scalasca analysis has been mainly carried out on the JUQUEEN system. Figs. 3-6 show the results of the Scalasca scaling analysis on JUQUEEN before (left panels) and after (right panels) optimisation of the code. Figs. 3 and 4 present the scaling of MPI calls and the non-MPI "USR" region of the code in comparison with the overall code ("ALL"). For strong scaling (Fig. 3) the speedup wrt. 1024 MPI tasks is shown, for weak scaling (Fig. 4) the overall time in seconds. Mind that the time spent in the "MPI" + "USR" + "COM" regions of the code is by definition equal to the total wall-clock time ("ALL"). Since the time spent in "COM" regions is negligible for our measurements, it is not shown in the scaling pictures for better readability. Figs. 5 and 6 show the scaling of the top 3 time consuming routines for strong and weak scaling, respectively. The instrumentation with Scalasca did not have any significant influence (typically below 5%) on the runtime behaviour and the total execution time of the program.

A Scalasca analysis of the initial version of the code shows that the core routines (i.e. everything but the I/O routines) scaled well on up to the largest tested configuration of 4096 MPI tasks for which the scaling performance is 97.2% and 100% of the ideal value for strong and weak scaling test, respectively. However, the scaling is heavily jeopardized by the I/O routines, which were requiring a major fraction of the wall-clock time when the number of MPI tasks was higher than 1024. One can clearly see from the left pictures of Figs. 3-6 that the behaviour of the MPI function *MPI_File_write* destroys the weak and strong scalability of the original code version. Most of the optimisation efforts were thus devoted to the new improved implementation of I/O (see section 3.1 for details). After a complete rewrite of the I/O scheme of the code, both the strong and the weak scaling behaviour have significantly improved as presented in the right pictures of Figs. 3-6. It was also possible now to test the code with up to 16384 MPI tasks.

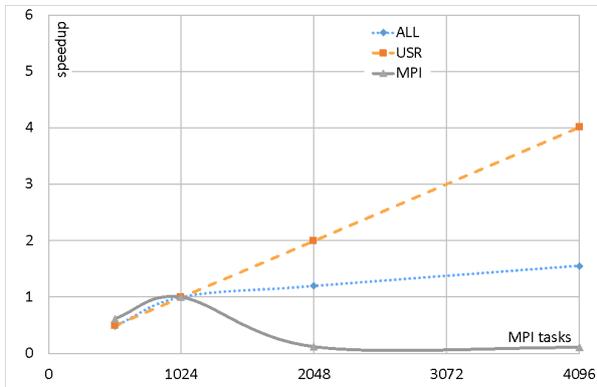 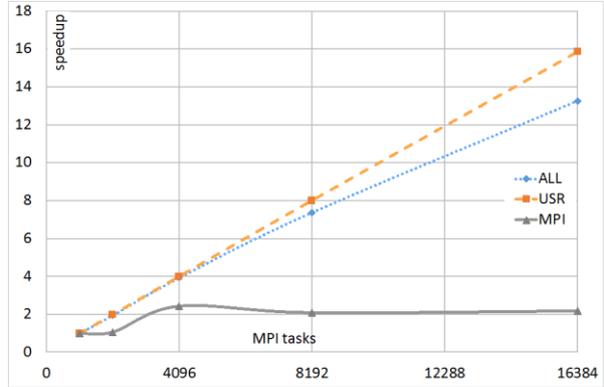

(a)            (b)

Figure 3 – Strong scaling of the "USR" and "MPI" parts of the 3D version of PICCANTE including I/O on JUQUEEN in comparison with the whole code ("ALL"). Fig. (a) shows the speedup wrt. 1024 MPI tasks before optimisation, Fig. (b) after optimisation of the I/O scheme.

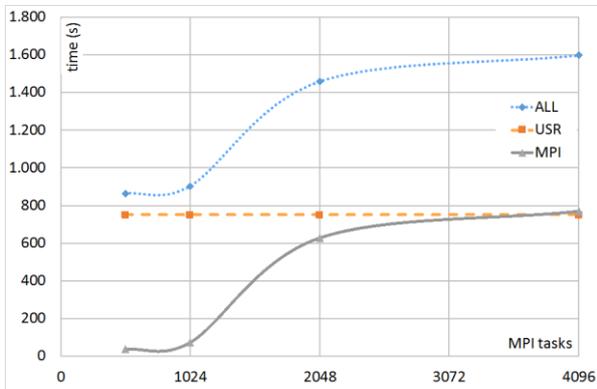 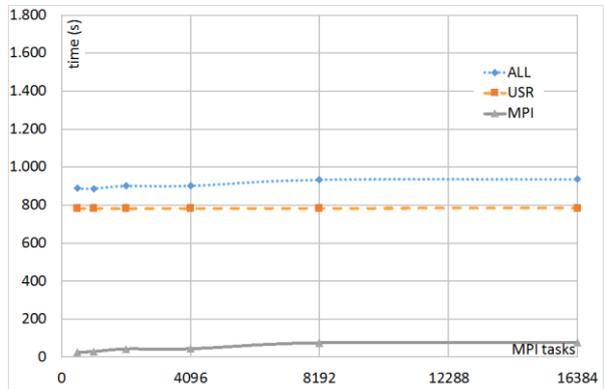

(a)            (b)

Figure 4 – Weak scaling of the "USR" and "MPI" parts of the 3D version of PICCANTE including I/O on JUQUEEN in comparison with the overall time ("ALL"). Fig. (a) shows the time before optimisation, Fig. (b) after optimisation of the I/O scheme

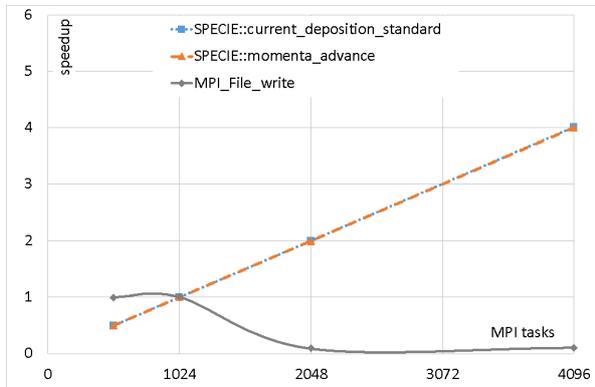 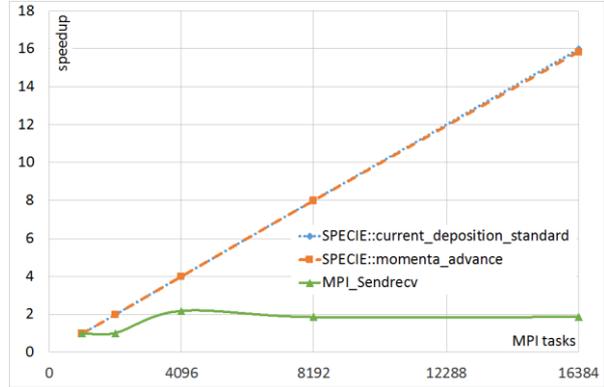

(a)                                                           (b)

Figure 5 – Strong scaling of the 3 most time consuming routines of the 3D version of PICCANTE including I/O on JUQUEEN. Fig. (a) shows the speedup wrt. 1024 MPI tasks of the most time consuming routines before optimisation, Fig. (b) after optimisation of the I/O scheme.

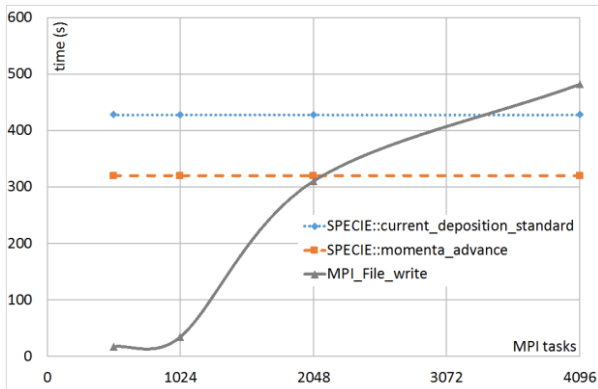 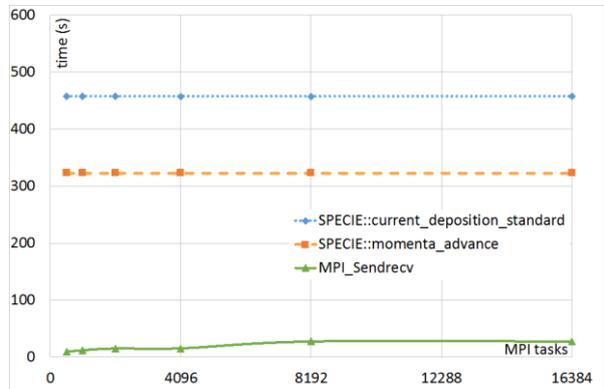

(a)                                                           (b)

Figure 6 – Weak scaling of the 3 most time consuming routines of the 3D version of PICCANTE including I/O on JUQUEEN. Fig. (a) shows the overall time spent in the most time consuming routines before optimisation, Fig. (b) after optimisation of the I/O scheme.

## 5. Summary and Outlook

The main goal of the project was to establish strong and weak scalability of the particle-in-cell code PICCANTE on large Tier-0 systems. Various optimisations have been performed. The complete rewriting of the I/O scheme had very good impact on the overall performance. All these developments have already been merged in the master branch of the repository available online at the project webpage [1]. The code is still under development, but a release tag was published for this whitepaper [13]. Further work will be concentrated on improving the hybrid version of the code using both OpenMP and MPI. The optimised version of the code is planned to be used in a regular PRACE project.

**Acknowledgements**

This work was financially supported by the PRACE project funded in part by the EUs 7th Framework Programme (FP7/2007-2015) under grant agreement no. RI-312763. The results were obtained within the PRACE Preparatory Access Type C Project 2010PA2458 "PICCANTE: an open source particle-in-cell code for advanced simulations on Tier-0 systems".